\documentclass[twocolumn,pre,aps]{revtex4-1}

\usepackage{graphicx}
\usepackage{epstopdf}
\usepackage{color}
\usepackage{footnote}
\usepackage{amsmath}
\usepackage{multirow}
\usepackage{graphicx}
\usepackage{ulem}

\begin{document}

\newcommand{\psen}{\pi}
\newcommand{\nsen}{\nu}
\newcommand{\sen}{\sigma}
\newcommand{\avsen}{\overline{\sigma}}
\newcommand{\BB}{\mathbf{B}}
\newcommand{\SSS}{\mathbf{S}}
\newcommand{\simi}{\sigma}
\newcommand{\s}{b}
\newcommand{\B}{B}

\renewcommand{\emph}{\textit}
\newcommand{\etal}{\textit{et al.}}
\newcommand{\etc}{\textit{etc.}}
\newcommand{\eref}[1]{(\ref{#1})}
\newcommand{\av}[1]{\langle#1\rangle} 
\newcommand{\bav}[1]{\bigl\langle#1\bigr\rangle} 
\newcommand{\bbav}[1]{\biggl\langle#1\biggr\rangle} 
\newcommand{\mean}[1]{\overline{#1}} 
\newcommand{\set}[1]{\{#1\}} 
\newcommand{\kw}[1]{\texttt{#1}}
\newcommand{\rats}{\vec{\pi}} 
\newcommand{\Ord}{\mathcal{O}} 
\newcommand{\ex}[1]{\mathrm{exp}(#1)} 
\newcommand{\defn}[1]{\textbf{#1}} 
\newcommand{\docs}{\mathcal{D}}
\newcommand{\words}{\mathcal{W}}
\newcommand{\topics}{\mathcal{T}}
\newcommand{\from}{\leftarrow} 
\newcommand{\half}{\frac{1}{2}} 
\newcommand{\third}{\frac{1}{3}} 
\newcommand{\twothirds}{\frac{2}{3}} 
\newcommand{\logten}{\log_{10}}

\newlength{\figurewidth}
\ifdim\columnwidth<10.5cm
  \setlength{\figurewidth}{0.95\columnwidth}
\else
  \setlength{\figurewidth}{10cm}
\fi
\setlength{\parskip}{0pt}
\setlength{\tabcolsep}{6pt}
\setlength{\arraycolsep}{2pt}

\title{Jury--Contestant Bipartite Competition Network: Identifying Biased Scores and Their Impact on Network Structure Inference}
\author{Gyuhyeon Jeon}
\affiliation{Graduate School of Culture Technology and BK21 Plus Postgraduate Program for Content Science, Korea Advanced Institute of Science \& Technology, Daejeon, Republic of Korea 34141}
\author{Juyong Park}
\affiliation{Graduate School of Culture Technology and BK21 Plus Postgraduate Program for Content Science, Korea Advanced Institute of Science \& Technology, Daejeon, Republic of Korea 34141}

\begin{abstract}
A common form of competition is one where judges grade contestants' performances which are then compiled to determine the final ranking of the contestants.  Unlike in another common form of competition where two contestants play a head-to-head match to produce a winner as in football or basketball, the objectivity of judges are prone to be questioned, potentially undermining the public's trust in the fairness of the competition.  In this work we show, by modeling the judge--contestant competition as a weighted bipartite network, how we can identify biased scores and measure their impact on our inference of the network structure. Analyzing the prestigious International Chopin Piano Competition of 2015 with a well-publicized scoring controversy as an example, we show that even a single statistically uncharacteristic score can be enough to gravely distort our inference of the community structure, demonstrating the importance of detecting and eliminating biases. In the process we also find that there does not exist a  significant system-wide bias of the judges based on the the race of the contestants.
\end{abstract}
\maketitle

\section{Introduction}
In a common form of competition, a group of judges scores contestants' performances to determine their ranking.  Unlike one-on-one pairwise direct competitions such as football or basketball where strict rules for scoring points must be followed and accordingly a clear winner is produced in the open, a complete reliance on the judges' subjective judgments can often lead to dissatisfaction by the fans and accusations of bias or even corruption~\cite{birnbaum1979source, zitzewitz2006nationalism,whissell1993national}.  Examples abound in history, including the Olympics that heavily feature the said type of competitions. A well-documented example is the figure skating judging scandal at the 2002 Salt Lake Olympics that can said to have been a prototypical judging controversy where the favorites lost under suspicious circumstances, which led to a  comprehensive reform in the scoring system~\cite{looney2003evaluating,kang2015dealing}.  A more recent, widely-publicized example can be found in the prestigious 17th International Chopin Piano Competition of 2015 in which judge Philippe Entremont gave contestant Seong-Jin Cho an ostensibly poor score compared with other judges and contestants. That Cho went on to win the competition nonetheless rendered the low score from Entremont all the more noteworthy, if not determinant of the final outcome~\cite{justin2015, onejuror2015}. Given that the competition format depends completely on human judgment, these incidents suggest that the following questions will persist: How do we detect a biased score? How much does a bias affect the outcome of the competition? What is the effect of the bias in our understanding of system's behavior?  Here we present a network framework to find answers and insights into these problems.

\begin{figure}
\resizebox{\figurewidth}{!}{\includegraphics{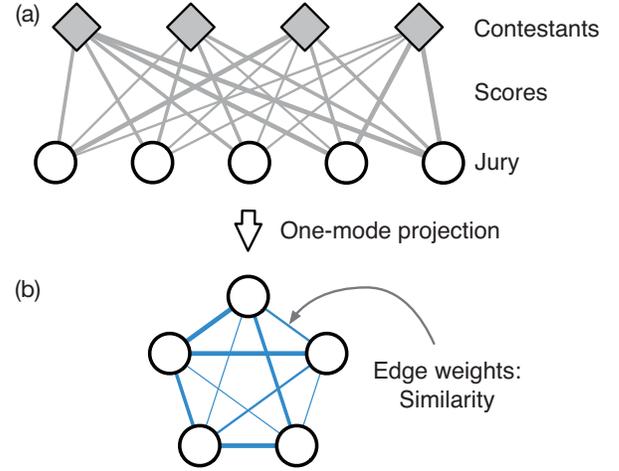}}
\caption{ (a) The bipartite network representation of the judge--contestant competition. The edge weight is the score given from a judge to a contestant.  (b) The one-mode projection onto the judges produces a weighted complete network whose edge weights are the similarity between the judges. We use the cosine similarity in our work.}
\label{definition}
\end{figure}

\begin{table*}[ht]
\caption{The final round scores from the 17th International Chopin Piano Competition. Judge Dang was not allowed to score the three contestants that were his former students (Liu, Lu, and Yang), noted ``NA''. The expected scores $\hat{b}$ of Eq.~\eref{expscore} in these cases are given in parentheses.}
\begin{tabular}{|c|c|c|c|c|c|c|c|c|c|c|}	
\hline
& Cho & Hamelin & Jurinic & Kobayashi & Liu & Lu & Osokins & Shiskin & Szymon & Yang \\ \hline

Alexeev & 10 & 8 & 2 & 1 & 7 & 9 & 3 & 6 & 4 & 5 \\ \hline
Argerich & 9 & 9 & 4 & 6 & 4 & 5 & 4 & 5 & 4 & 5 \\ \hline
Dang & 8 & 8 & 2 & 7 & \begin{tabular}{@{}c@{}}NA \\ (6.124) \end{tabular} & \begin{tabular}{@{}c@{}}NA \\ (5.673) \end{tabular} & 1 & 5 & 4 & \begin{tabular}{@{}c@{}}NA \\ (4.776) \end{tabular}  \\ \hline
Ebi & 9 & 9 & 3 & 5 & 3 & 4 & 5 & 5 & 3 & 8 \\ \hline
Entremont & 1 & 8 & 3 & 2 & 5 & 8 & 4 & 7 & 4 & 6 \\ \hline
Goerner & 9 & 10 & 2 & 5 & 5 & 8 & 2 & 6 & 2 & 6 \\ \hline
Harasiewicz & 6 & 7 & 6 & 2 & 9 & 3 & 5 & 7 & 2 & 2 \\ \hline
Jasiński & 9 & 6 & 3 & 8 & 10 & 6 & 2 & 3 & 5 & 2 \\ \hline
Ohlsson & 9 & 8 & 6 & 1 & 9 & 4 & 5 & 7 & 2 & 3 \\ \hline
Olejniczak & 10 & 7 & 1 & 5 & 9 & 8 & 3 & 2 & 6 & 4 \\ \hline
Paleczny & 9 & 6 & 1 & 4 & 10 & 8 & 2 & 3 & 5 & 7 \\ \hline
Pobłocka & 9 & 7 & 1 & 6 & 8 & 8 & 2 & 5 & 2 & 6 \\ \hline
Popowa-Zydroń & 9 & 10 & 1 & 6 & 9 & 8 & 1 & 1 & 4 & 6 \\ \hline
Rink & 9 & 9 & 5 & 3 & 8 & 4 & 7 & 6 & 2 & 1 \\ \hline
Świtała & 9 & 8 & 1 & 5 & 10 & 7 & 4 & 1 & 5 & 5 \\ \hline
Yoffe & 9 & 9 & 5 & 3 & 8 & 7 & 6 & 2 & 4 & 2 \\ \hline
Yundi & 9 & 9 & 4 & 5 & 6 & 6 & 2 & 4 & 3 & 5 \\ \hline
\end{tabular}
\label{BB}
\end{table*}

\section{Methodology and Analysis}
The judge--contestant competition can be modeled as a bipartite network with weighted edges representing the scores, shown in Fig.~\ref{definition}~(a).  It is a graphical representation of the $l\times r$--bipartite adjacency matrix
\begin{align}
	\BB = \set{b_{ij}},
\end{align}
whose actual values from the 17th International Chopin Piano Competition are given in Table~\ref{BB}~\cite{finaloceny2015}, composed of $l=17$ judges and $r=10$ contestants. The entries ``NA'' refer to the cases of judge Thai Son Dang ($i=3$) and his former pupils Kate Liu ($j=4$), Eric Lu ($j=5$), and Yike (Tony) Yang ($j=10$) whom he was not allowed to score. For convenience in our later analysis, we nevertheless fill these entries with expected scores based on their scoring tendencies using the formula
\begin{align}
	\hat{b}_{ij} = \sqrt{\frac{\sum'_ib_{ij}}{l-1}\times\frac{\sum''_jb_{ij}}{r-3}},
\label{expscore}
\end{align}
where the summations $\sum'$ and $\sum''$ indicate omitting these individuals. It is the geometric mean of Dang's average score given to the other contestants and the contestant's average score obtained from the other judges. The values are given inside parentheses in Table~\ref{BB}.  A one-mode projection of the original bipartite network onto the judges is shown in Fig.~\ref{definition}~(b), which is also weighted~\cite{zhou2007bipartite}. The edge weights here indicate the similarity between the judges, for which we use the cosine similarity
\begin{align}
	\simi_{ij} \equiv \frac{\vec{\s}_i\cdot\vec{\s}_j}{|\vec{\s}_i||\vec{\s}_j|} = \frac{\sum_kb_{ik}b_{jk}}{\sqrt{\sum_kb_{ik}^2}\sqrt{\sum_kb_{ik}^2}}
\label{cosine}
\end{align}
with $\hat{b}$ naturally substituted for $b$ when applicable. This defines the $17\times 17$ adjacency matrix $\SSS=\set{\simi_{ij}}$ of the one-mode projection network.

\begin{figure}
\resizebox{\figurewidth}{!}{\includegraphics{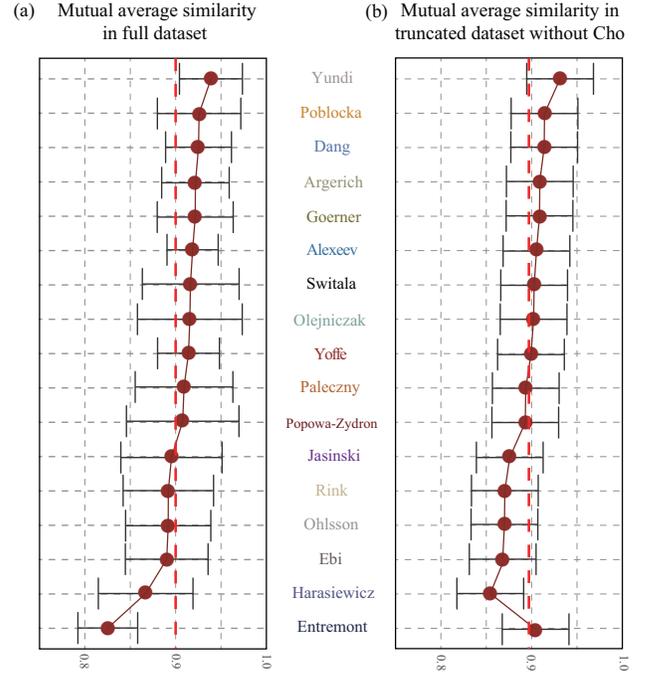}}
\caption{The mutual average similarity between the judges. (a) In the full data Philippe Entremont is the most dissimilar, atypical judge. (b) In the truncated data with Cho removed, Entremont now becomes the seventh most similar judge, placing himself as the more typical one. This abrupt change indicates the outsize effect of the single uncharacteristic score given to Cho by Entremont.}
\label{meansimilarity}
\end{figure}

\subsection{Determining the Atypicality of Judges}
Perhaps the most straightforward method for determining the atypicality of a judge before analyzing the network (i.e., using the full matrix) is to compare the judges' average mutual similarities (average similarity to the other judges), shown in Fig.~\ref{meansimilarity}.  Using the full data set (Fig.~\ref{meansimilarity}~(a)) we find Yundi to be the most similar to the others with $\av{\sigma}=0.94\pm0.04$, and Entremont to be the least so with $\av{\sigma}=0.83\pm 0.03$. The global average similarity is $0.90\pm 0.05$, indicated by the red dotted line.  Yundi is not particularly interesting for our purposes, since a high overall similarity indicates that he is the most typical, average judge.  Entremont, on the other hand, is the most interesting case.  Given the attention he received for his low score to Cho, this makes us wonder how much of this atypicality of his was a result of it.  To see this, we perform the same analysis with Cho removed from the data, shown in Fig.~\ref{meansimilarity}~(b).  Entremont is now ranked 7th in similarity, indicating that his score on Cho likely was a very strong factor for his atypicality first seen in Fig.~\ref{meansimilarity}~(a).

\begin{figure*}[htpb]
  \begin{center}
  	\includegraphics[scale=0.5]{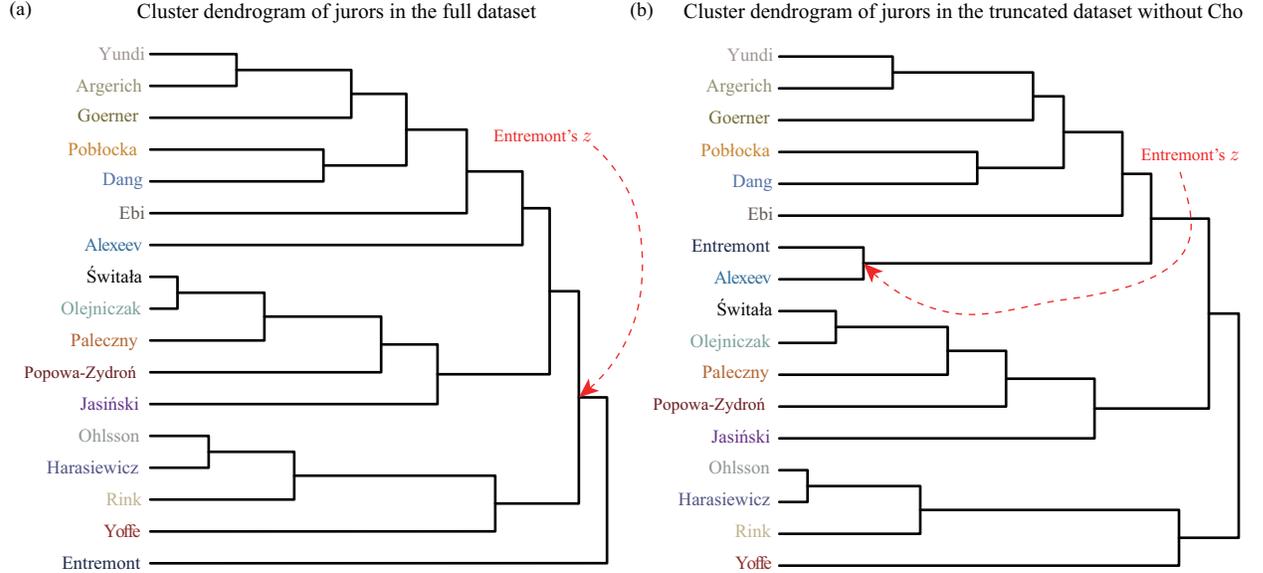}
	\caption {Clustering dendrogram of the judges in 17th Chopin Competition. (a) Using the full data. Entremont joins the dendrogram at the last possible level ($z=16$). (b) Using the truncated data with Cho removed. Entremont joins the dendrogram early, at level $z=3$. The elimination of a single contestant has rendered Entremont to appear to be one of the more typical judges, consistent with Fig.~\ref{meansimilarity}~(b).}
	\label{clusterDendro}
  \end{center}
  \end{figure*}

Although the effect of a single uncharacteristic score was somewhat demonstrated in Fig.~\ref{meansimilarity}, by averaging out the edge weights we have incurred a nearly complete loss of information on the network structure.   We now directly study the network and investigate the degree to which biased scores affect its structure and our inferences about it.  While there exists a wide variety of analytical and computational methods for network analysis~\cite{newman2010networks,barabasi201609}, here we specifically utilize \emph{hierarchical clustering} for exploring our questions at hand.  Hierarchical clustering is most often used in classification problems by identifying clusters or groups of objects based on similarity or affinity between them~\cite{johnson1967hierarchical,newman2004fast,murtagh2012algorithms}.  The method's name contains the word ``hierarchical'' because it produces a hierarchy of groups of objects starting from each object being its own group at the bottom to a single, all-encompassing group at the top.  The hierarchy thus found is visually represented using a dendrogram such as the one shown in Fig.~\ref{clusterDendro}, generated for the judges based on cosine similarity $\simi_{ij}$ of Eq.~\eref{cosine}.  We used agglomerative clustering with average linkage~\cite{eisen1998cluster,newman2004finding}.  Before we use the dendrogram to identify clusters, we first focus on another observable from the dendrogram, the level $z$ at which a given node joins the dendrogram. A node with small $z$ joins the dendrogram early, meaning a high level of similarity with others; a large $z$ means the opposite. This $z$ is consistent with Fig.~\ref{meansimilarity}: For Entremont $z=16$ (the maximum possible value with 17 judges) in the full data set, being the last one to join the dendrogram in the full data set, while $z=3$ when Cho is removed.  The two dendrograms and Entremont's $z$ are shown in Fig.~\ref{clusterDendro}~(a)~and~(b).  $z$ is therefore a  simple and useful quantity for characterizing a node's atypicality.  To see if any other contestant had a similar relationship with Entremont, we repeat this process by removing the contestants alternately from the data and measuring Entremont's $z$, the results of which are shown in Fig.~\ref{entryPoint}.  No other contestant had a similar effect on Entremont's $z$, once again affirming the uncharacteristic nature of Entremont's score of Cho's performance.

\subsection{Racism as a Factor in Scoring}
\begin{figure}
	\resizebox{\figurewidth}{!}{\includegraphics{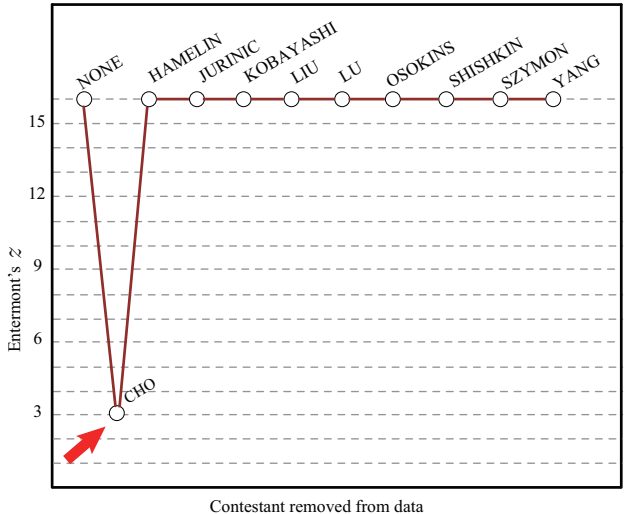}}
	\caption{Entremont's entry point $z$ into the dendrogram as the contestants are alternately removed from the data one by one. The removal of any contestant other than Cho has no visible effect on the typicality of of Entremont.}
	\label{entryPoint}
\end{figure}

\begin{figure*}[htpb]
  \begin{center}
  	\includegraphics[scale=0.5]{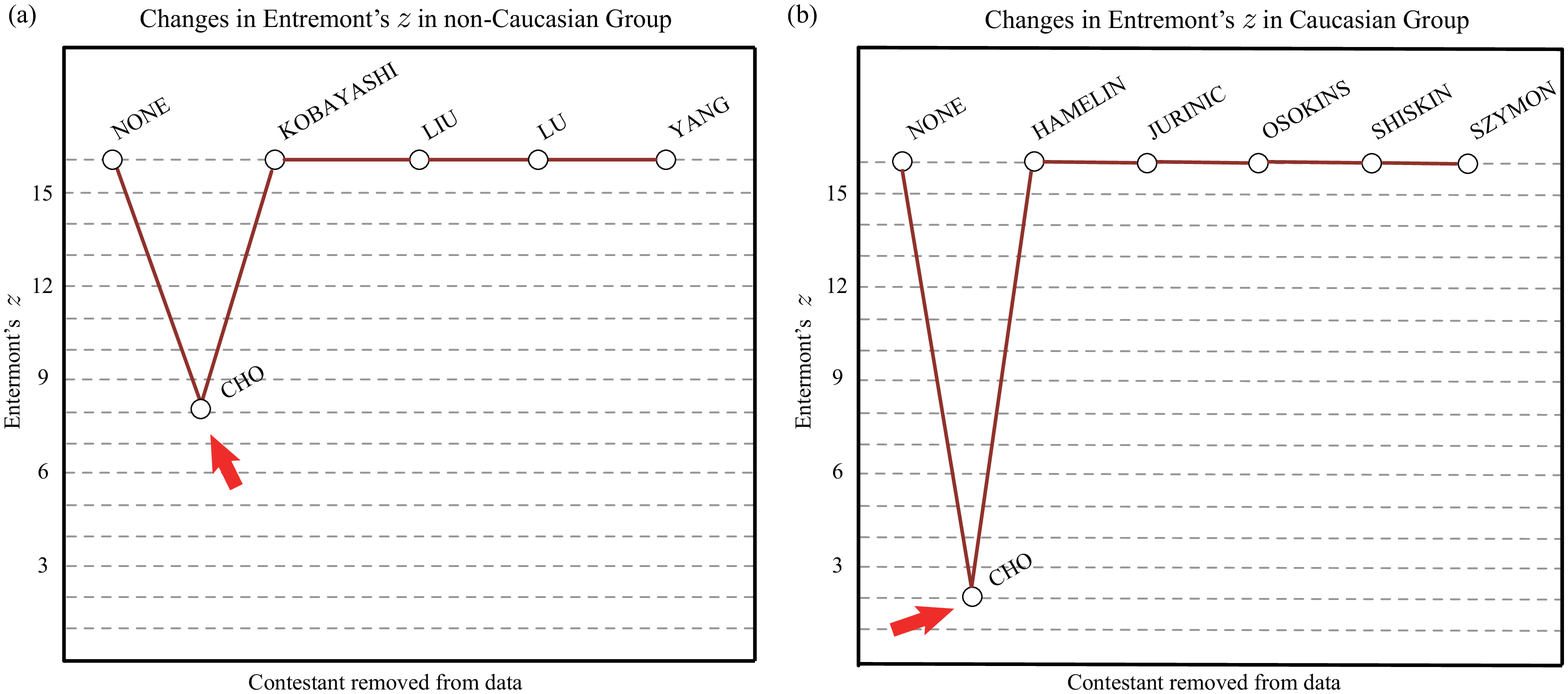}
	\caption {Entremont's entry point into the clustering dendrogram in two distinct groups of contestants, the non-Caucasian group (a) and the Caucasian group plus Cho (b). In both cases, Entremont's $z$ decreases from the maximum value only when Cho is removed. This suggests that Entremont's uncharacteristic score of Cho's performance was not ethnically motivated.}
	\label{R5_EastWestEntryPoint}
  \end{center}
  \end{figure*}

A popular conjecture regarding the origin of Cho's low score was that Entremont may have been racially motived. We can perform a similar analysis to find any such bias against a specific group (e.g., non-Caucasians) of contestants. To do so, we split the contestants into two groups, non-Caucasians and Caucasians plus Cho (the ethnicities were inferred from their surnames and, when available, photos) as follows:
\begin{itemize}
	\item Non-Caucasians (5): Cho, Kobayashi, Liu, Lu, and Yang
	\item Caucasians plus Cho (6): Cho, Hamelin, Jurinic, Osokins, Shiskin, and Szymon.
\end{itemize}
We then plot figures similar to Fig.~\ref{entryPoint}. If Entremont had truly treated the two racial groups differently, the effect of his score to Cho would have had significantly different effects on each group.  The results are shown in Fig.~\ref{R5_EastWestEntryPoint}.  As before, Entrement's low score of Cho stands out amongst the non-Caucasian contestants, a strong indication that the race hadn't played a role, although it should be noted that Entremont appears to have been more dissimilar overall from the other judges in scoring the non-Caucasian contestants when Cho was not considered ($z=8$ compared with $z=2$ in the Caucasian group).

\subsection{Impact of Biased Edges on Inference of Network's Modular Structure}
As pointed out earlier, hierarchical clustering is most often used to determine the modular structure of a network. This is often achieved by making a ``cut'' in the dendrogram on a certain level~\cite{langfelder2008defining}.  A classical method for deciding the position of the cut is to maximize the so-called \emph{modularity} $Q$ defined as
\begin{align}
	Q = \frac{1}{2m}\sum_{ij}\biggl[A_{ij}-\frac{k_ik_j}{2m}\biggr]\delta(c_i,c_j)
\label{modularity}
\end{align}
where $m$ is the number of edges, $c_i$ is the module that node $i$ belongs to, and $\delta$ is the Kronecker delta~\cite{newman2006modularity}.  The first factor in the summand is the difference between the actual number of edges ($0$ or $1$ in a simple graph) between a node pair and its random expectation based on the nodes' degrees.   We now try to generalize this quantity for our one-mode projection network in Fig.~\ref{definition}~(b) where the edge values are the pairwise similarities $\simi\in[0,1]$. At first a straightforward generalization of Eq.~\eref{modularity} appears to be, disregarding the $(2m)^{-1}$ which is a mere constant, 
\begin{align}
	Q' = \sum_{ij}\bigl(\simi_{ij}-\av{\simi_{ij}}\bigr)\delta(c_i,c_j)\equiv\sum_{ij}q'_{ij}\delta(c_i,c_j)
\label{qtemp}
\end{align}
where $\av{\simi_{ij}}$ is the expected similarity obtained by randomly shuffling the scores (edge weights) in the bipartite network of Fig.~\ref{definition}~(a).  In the case of the cosine similarity this value can be computed analytically using its definition Eq.~\eref{cosine}: it is equal to the average over all permutations of the elements of $\vec{\s}_i$ and $\vec{\s}_j$.  What makes it even simpler is that permutating either one is sufficient, say $\vec{\s}_j$.  Denoting by $\vec{\s}_j^{(k)}$ the $k$-th permutation of $\vec{\s}$ out of the $r!$ possible ones, we have
\begin{align}
	\av{\simi_{ij}}
	&= \frac{1}{r!|\vec{\s}_i||\vec{\s}_j|} \biggl(\vec{\s}_i\cdot\vec{\s}_j^{(1)}+\cdots+\vec{\s}_i\cdot\vec{\s}_j^{(r!)}\biggr) \nonumber \\
	&= \frac{1}{r!|\vec{\s}_i||\vec{\s}_j|}\vec{\s}_i\cdot\biggl(\vec{\s}_j^{(1)}+\cdots+\vec{\s}_j^{(r!)}\biggr)\nonumber \\
	&= \frac{1}{r!|\vec{\s}_i||\vec{\s}_j|}\biggl(\s_{i1}\bigl(\s_{j1}^{(1)}+\cdots+\s_{j1}^{(r!)}\bigr) \nonumber \\
	&~~~+\s_{i2}\bigl(\s_{j2}^{(1)}+\cdots+\s_{j2}^{(r!)}\bigr)+\cdots\bigr)\nonumber \\
	&=\frac{(\sum_{k=1}^r\s_{ik})(\sum_{k=1}^r\s_{jk})}{r!|\vec{\s}_i||\vec{\s}_j|}\equiv\frac{\B_i\B_j}{r!|\vec{\s}_i||\vec{\s}_j|}.
\label{expectation}
\end{align}

When we insert this value into Eq.~\eref{qtemp} and try to maximize it for our network, however, we end up with a single module that contains all the judges as the optimal solution, a rather uninteresting and uninformative result.  On closer inspection, in turns out, this stems from the specific nature of summand $q'$ with regards to our network.  For a majority of node pairs the summand is positive (even when Entremont is involved), so that it is advantageous to have $\delta(c_i,c_j)=1$ for all $(i,j)$ for $Q'$ to be positive and large, i.e. all judges belonging to a single, all-encompassing module, as noted. The reason why $Q$ of Eq.~\ref{modularity} has worked so well for sparse simple networks was that most summands were negative (since $A_{ij}=0$ for most node pairs in a sparse network, and $k_ik_j/(2m)>0$ always), so that including all nodes in a single group was not an optimal solution for $Q$.  To find a level of differentiation between the judges, therefore, we need to further modify $Q'$ so that we have a reasonable number of negative as well as positive summands. We achieve this by subtracting a universal positive value $q'_0$ from each summand, which we propose to be the mean of the $q'$, i.e.
\begin{align}
	q'_0 &= \mean{q'}=\mean{\simi_{ij}-\av{\simi_{ij}}}
		= \frac{2}{l(l-1)}\sum_{i<j}\bigl(\simi_{ij}-\av{\simi_{ij}}\bigr) \nonumber \\		
		&=\mean{\sigma}-\mean{\av{\sigma}}.
\end{align}
At this point one must take caution not to be confused by the notations: $\mean{\simi}$ is the average of the actual similarities from data, while $\av{\simi}$ is the pairwise random expectation from Eq.~\eref{expectation}, and therefore $\mean{\av{\sigma}}$ is the average of the pairwise random expectations. Then our re-modified modularity is
\begin{align}
	Q''	&\equiv\sum_{ij}q_{ij}''\delta(c_i,c_j) \equiv \sum_{ij}\bigl(q'_{ij}-q'_0\bigr)\delta(c_i,c_j)\nonumber \\
		&=\sum_{ij}\bigl[(\simi_{ij}-\av{\simi_{ij}})-\overline{(\simi_{ij}-\av{\simi_{ij}})}\bigr]\delta(c_i,c_j)
\end{align}
This also has the useful property of vanishing to $0$ at the two ends of the dendrogram (i.e. all nodes being separate or forming a single module), allowing us to naturally avoid the most trivial or uninformative cases.

\begin{figure*}[htpb]
  \begin{center}
  	\includegraphics[scale=0.75]{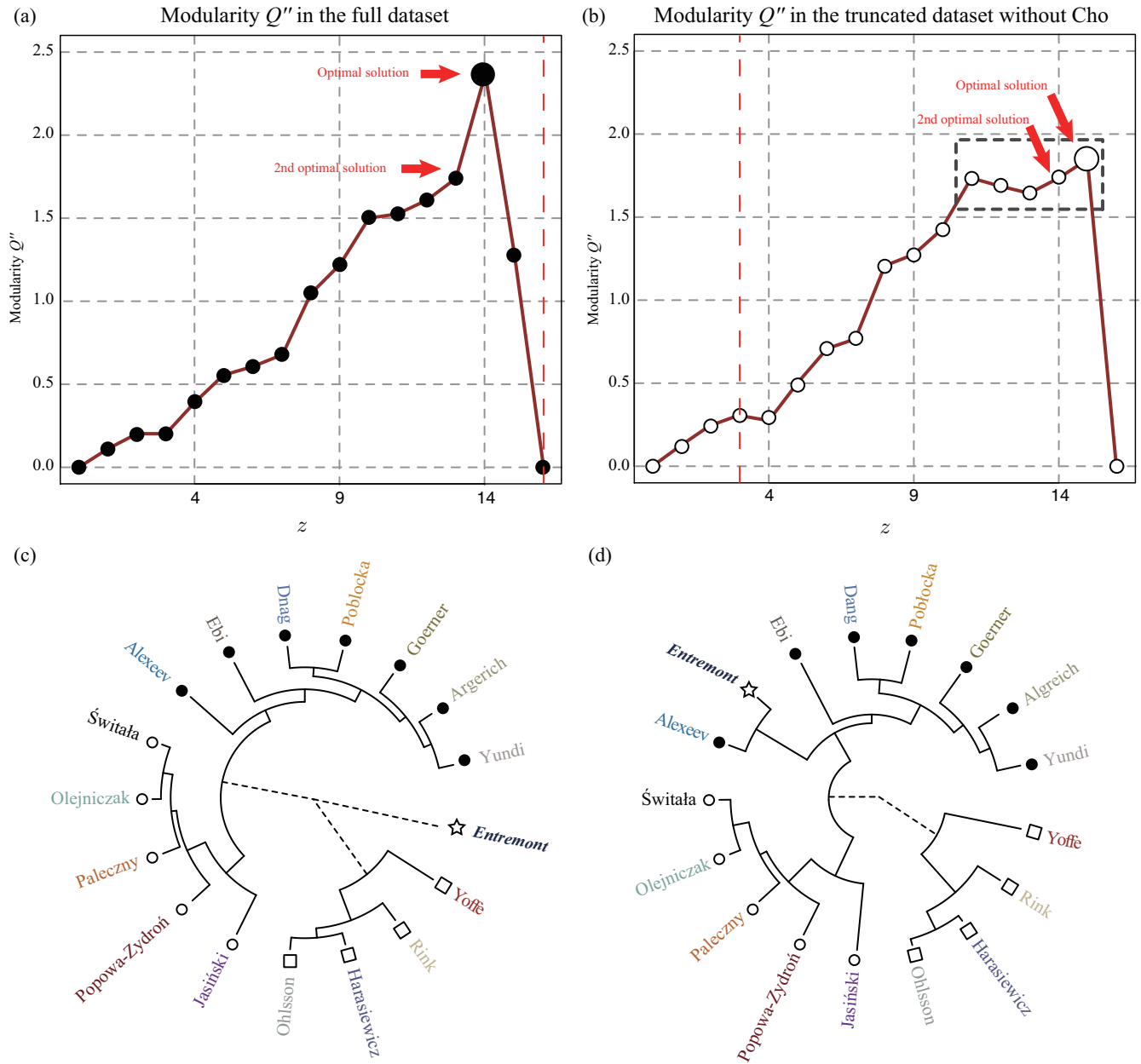}
	\caption {Modified modularity $Q''$ and the modular structure of the judges' network. (a) In the full network, maximum $Q''$ occurs at $z=14$, resulting in three modules shown in (c). (b) In the truncated network with Cho removed the optimal solution is at $z=15$, resulting in two modules shown in (d). A comparison of (c) and (d) tells us that the single biased score can be powerful enough to produce cause Entremont to form his own module, and also present the solution as the dominant one; in the absence of the edge, several comparable solutions can exist (marked by an encompassing rectangle) in (b).}
	\label{clusterNetworks}
  \end{center}
\end{figure*}

We now plot $Q''$ as we traverse up the dendrograms from Figs.~\ref{clusterNetworks}. For the full data with Cho, maximum $Q''$ occurs at level $z=14$, yielding $3$ modules (Fig.~\ref{clusterNetworks}~(c)).  With Cho removed, in contrast, maximum $Q''$ occurs at level $z=15$, resulting in $2$ modules (Fig.~\ref{clusterNetworks}~(d)).  In the full data Entremont forms an isolated module on his own, but otherwise the $Q''$-maximal modular structures are identical in both cases. This is another example of how a single uncharacteristic, biased score from Entremont to Cho is responsible for a qualitatively different observed behavior of the system.  There is another issue that warrants further attention, demonstrating the potential harm brought on by a single biased edge:  We see in Fig.~\ref{clusterNetworks}~(a) that the $Q''$-maximal solution ($z=14$) eclipses all the possibilities ($\Delta Q''=2.3687$ between it and the second optimal solution, for a relative difference $\Delta Q''/Q''_{\textrm{max}}=0.2653$), compared with Fig.~\ref{clusterNetworks}~(b) where the difference between the two most optimal solutions is much smaller ($\Delta Q''=1.8430$ and $\Delta Q''/Q''_{\textrm{max}}=0.0554$). Furthermore, there are at least three other solutions with comparable $Q''$ $(z=13,~12,~11)$ in Fig.~\ref{clusterNetworks}~(b). Given the small differences in $Q''$ between these solutions, it is plausible that had we used slightly a different definition of modularity or tried alternative clustering methods, any of these or another comparable configuration may have presented itself as the optimal solution.  But a single biased edge was so impactful that not only an apparently incorrect solution was identified as the most optimal, but also much more dominant than any other.

\section{Discussions and Conclusions}
Given the prevalence of competition in nature and society, it is important to understand the behaviors of different competition formats know their strengths, weakness, and improve their credibility.  Direct one-on-one competitions are the easiest to visualize and model as a network, and many centralities can be applied either directly or in a modified form to produce reliable rankings~\cite{park2005network, shin2014ranking}.  Such competition formats are mostly free from systematic biases, since the scores are direct results of one competitor's superiority over the other. The jury--contestant competition format, while commonly used, provides a more serious challenge since it relies completely on human judgement; when the public senses unwarranted bias they may lose trust in the fairness of the system, which is the most serious threat against the very existence of a competition.

Here we presented a network study of the jury-contestant competition, and showed how we can use the hierarchical clustering method to detect biased scores and measure their impact on the network structure.   We began by first identifying the most abnormal jury member in the network, i.e. the one that is the least similar. While using the individual jury member's mean similarity to the others had some uses, using the dendrogram to determine the atypicality of a judge graphically was more intuitive and allowed us gain a more complete understanding of the network. After confirming the existence of a biased score, we investigated the effect of the bias on the network structure.  For this analysis, we introduced a modified modularity measure appropriate for our type of network.  This analysis revealed in quite stark terms the dangers posed by such biased edges; even a single biased edge that accounted for less than 1\% of the edges led us to make unreliable and misleading inferences about the network structure. 

Given the increasing adoption of the network framework for data modeling and analysis in competition systems where fairness and robustness are important, we hope that our work highlights the importance of detecting biases and understanding their effect on network structure.

\acknowledgements{
This work was supported National Research Foundation of Korea (NRF-20100004910 and NRF-2013S1A3A2055285), BK21 Plus Postgraduate Organization for Content Science, and the Digital Contents Research and Development program of MSIP (R0184-15-1037, Development of Data Mining Core Technologies for Real-time Intelligent Information Recommendation in Smart Spaces)
}

\bibliographystyle{apsrev4-1}
\bibliography{Chopin}

\end{document}